\documentclass[12pt]{iopart}
\usepackage{graphicx}
\usepackage{amssymb}
\usepackage[latin1]{inputenc}
\usepackage{color}
\begin{document}

\title{Reduction of heating rate in a microfabricated ion trap by pulsed-laser cleaning}

\author{D T C Allcock$^1$, L Guidoni$^{1,2}$, T P Harty$^1$,  C J Ballance$^1$, M G Blain$^3$, A M Steane$^1$ and D M Lucas$^1$}
\address{$^1$ Department of Physics, University of Oxford, Clarendon Laboratory, Parks Road, Oxford OX1 3PU, UK}
\address{$^2$ Univ. Paris Diderot, Sorbonne Paris Cité, Laboratoire Matériaux et Phénomènes Quantiques, UMR 7162 CNRS, F-75205 Paris, France.}
\address{$^3$ Sandia National Laboratories, Albuquerque, New Mexico 87185, USA}
\ead{d.allcock@physics.ox.ac.uk}

\begin{abstract}
Laser-cleaning of the electrodes in a planar micro-fabricated ion trap has been attempted using ns pulses from a tripled Nd:YAG laser at 355nm.
The effect of the laser pulses at several energy density levels has been tested by measuring 
the heating rate of a single $^{40}$Ca$^+$ trapped ion as a function of its secular frequency $\omega_z$.
A reduction of the electric-field noise spectral density by $\sim50\%$ has been observed and a change in the frequency dependence also noticed.
This is the first reported experiment where the ``anomalous heating'' phenomenon has been reduced by removing the source as opposed to reducing its thermal driving by cryogenic cooling.
This technique may open the way to better control of the electrode surface quality in ion microtraps.

\end{abstract}

%\pacs{00.00}
%\submitto{\NJP}
\maketitle

%\section{Introduction}
%\label{sec:intro}
The recent success of quantum information experiments based on trapped ions~\cite{Haffner:2008} triggered research on micro-fabricated radio-frequency (Paul) traps, which are in principle able to fulfill the scalability requirement of a quantum computer~\cite{DiVincenzo:2000, Steane:2007}.
In such traps, a set of micro-fabricated conducting electrodes generates oscillating and static electric fields that trap laser-cooled ions in a harmonic potential well at a sub-millimeter distance $d$ from the substrate~\cite{Chiaverini:2005, Seidelin:2006}.
However, the presence of uncontrolled fluctuating electric fields affects the ions' external motion and induces an ``anomalous heating'' that limits the achievable fidelity of multi-ion quantum gates that rely on the coherent control of this motion~\cite{Wineland:1998}.
Experimental observations concerning this phenomenon are consistent with a very unfavorable $d^{-4}$ scaling law, compatible with a random distribution of fluctuating charges or dipolar ``patches'' at the electrode surfaces~\cite{Turchette:2000, Deslauriers:2004}.
In addition, the scaling of the electric field noise spectral density $S_E(\omega)$ with respect to trap secular frequency $\omega$ has been found to approximate a $\omega^{-\alpha}$ law, with exponents roughly compatible with $\alpha=1$ but spanning the range $0.4 < \alpha<1.6$\quad\cite{Turchette:2000,Deslauriers:2004,Deslauriers:2006,Epstein:2007,Labaziewicz:2008a,Allcock:2011}.

Some recent theoretical models propose the fluctuations of the electric dipoles of adsorbed molecules as a possible driving mechanism~\cite{Daniilidis:2011,Safavi-Naini:2011}, whilst other authors point out the role played by a more general (but microscopically not identified) correlation length associated with disorder on the surface~\cite{Dubessy:2009,Low:2011}.
Several studies point to surface contamination being an issue: the first observed an order of magnitude variation between four nominally identical traps and even the same trap after it had been re-cleaned~\cite{Labaziewicz:2008a}, the second observed no change in the heating rate even when the bulk of the electrode undergoes a transition to a superconducting state~\cite{Wang:2010}, and the third reported an increased heating rate in the region of the trap used for loading ions~\cite{Daniilidis:2011}.
Although traps are typically cleaned using some combination of organic solvents, ozone cleaning or plasma cleaning after fabrication, any surface that has been exposed to atmosphere will have a covering of adsorbents at least several mono-layers thick.
Additionally, trap electrode materials that react with oxygen will have a native oxide layer.
Standard methods of preparing atomically clean metal surfaces under ultra-high vacuum involve either {\it in situ} cleaving, evaporation, or repeated cycles of ion bombardment and high temperature annealing~\cite{Woodruff:1994}.
Whilst the latter two processes could be used in principle for microtraps, they would add significant engineering complexity to the trap structures and to the experimental vacuum systems.
Moreover, such harsh treatments are unlikely to be compatible with traps currently under development which incorporate integrated optics~\cite{Streed:2011,Brady:2011}, an important step towards scaling up ion trap quantum computing. 

%
%%%%%%%%
 \begin{figure}[h]
   \centerline{\includegraphics[width=.75\columnwidth]{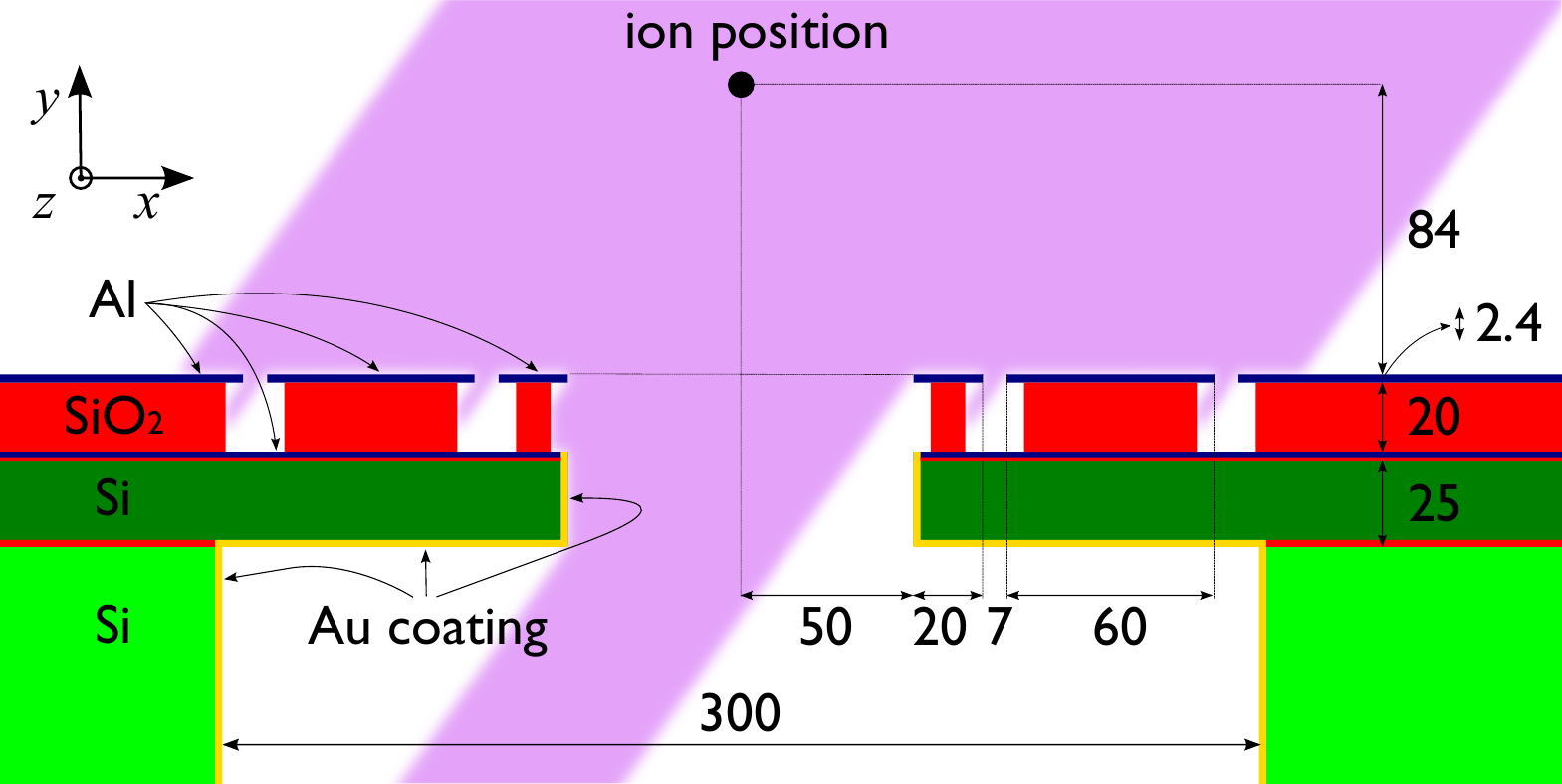}}
  \caption{Schematic cross-section (to scale) of the microfabricated trap~\cite{Allcock:2011,Stick:2010}; all dimensions are in microns. The slot through the centre of the substrate, parallel to the trap $z$ axis, allows ions to be loaded from the underside of the substrate: this avoids contamination of the trap electrodes from the neutral atomic beam source. An important issue for the current study is the gold coating that covers the silicon, which has a nominal thickness of 0.11~$\mu$m on the slot side walls (not to scale on the scheme). The cleaning beam cross-section corresponding to the $\vec{k_-}=(-\frac{1}{2},-\frac{1}{\sqrt{2}},\frac{1}{2})$ propagation direction (violet shade) is also sketched.}
  \label{fig:scheme_trap}
\end{figure}
%%%%%%%%
%
The cleaning of metallic surfaces based on pulsed-laser sources has also been noted as an effective, if less frequently used, technique for producing clean surfaces~\cite{Delaporte:2006}.
The technique is based on the fact that energy density thresholds for desorbing contaminants or removing oxides are generally lower than the ablation damage threshold for the metallic surface~\cite{Phipps:2007}.
In particular, ``dry'' laser cleaning, compatible with ultra-high vacuum techniques, has been used for oxide removal from metallic surfaces~\cite{Meja:1999,Lee:2000} and cleaning of aluminium-coated optical surfaces~\cite{Mann:1996}.
Typically, ultraviolet pulses from nanosecond sources (e.g. excimer or Nd:YAG 3rd/4th harmonic) and energy densities of $\sim 100$~mJ/cm$^2$ are used.
Laser-cleaning may be easily applied to an ion microtrap and it requires no modification to a typical vacuum system (viewports which transmit UV are often required for laser access).
Furthermore, the cleaning-laser beam can be positioned with micron-level precision and its direction easily adjusted, in order to avoid delicate components or to irradiate  selectively different parts of complex 3D trap designs.

%\section{Experimental methods}
%\label{sec:exp}
We implement this technique on a state-of-the-art microfabricated trap~\cite{Allcock:2011,Brady:2011,Stick:2010}.
The structure of the trap (see figure~\ref{fig:scheme_trap}) is such that three different materials are exposed to the cleaning beam: the aluminium of the upper electrode surface (2.4~$\mu$m of sputter deposited Al--1/2\%~Cu with 2--3~nm native oxide having an RMS surface roughness of $\sim 8$~nm), the gold coating on the silicon (500~nm Au / 50 nm Pt / 20 nm Ti stack, e-beam evaporated) and the silicon dioxide of the pillars which support the electrodes (plasma deposited TEOS).
We note that the gold coating was evaporated at an angle such that it has a nominal thickness of 114~nm on the slot side walls.
Previous works describe laser ablation or laser cleaning of such materials; we briefly review here the main results that may apply to the present study.
For aluminium the generally accepted ablation threshold (for plasma generation) lies around 4~J/cm$^2$ at $\lambda=355$~nm~\cite{Petzoldt:1996}; however, careful studies in high vacuum demonstrated that a measurable Al$^+$ ion yield appears at energy densities lower than 100~mJ/cm$^2$~\cite{Taylor:2000}.
At the same wavelength the reported thresholds for cleaning and damaging the surface of an Al coated glass substrate (BK7)  are 200~mJ/cm$^2$  and 490~mJ/cm$^2$ respectively~\cite{Mann:1996}.
The laser ablation of aluminium oxide is somewhat more complicated due to the fact that several phases can coexist in native oxides.
An experiment performed in ultra-high vacuum conditions (ion detection and surface analysis) on a sapphire monocrystal~\cite{Schildbach:1992} gives an ablation threshold of 3~J/cm$^2$ and a threshold more than one order of magnitude lower for Al$^+$ ion emission.
Even lower thresholds are expected for native oxides~\cite{Meja:1999}.
In the case of gold, an experimental study addressed the case of thin films (up to some microns) in the single-shot regime~\cite{Matthias:1994} and estimated the ablation threshold to be $\sim 250$~mJ/cm$^2$ (with the damage threshold a factor of 2 below this) for a film thickness of 100~nm.
While the ablation threshold for silicon is well-documented (1.3~J/cm$^2$)~\cite{Gusev:1995}, the case of silicon dioxide is less straightforward to analyze, due to the differences in composition and porosity.

%
%%%%%%%%
 \begin{figure}[h]
%   \centerline{\hfil \includegraphics[width=.45\columnwidth]{fig2Abis} \hfil \includegraphics[width=.45\columnwidth]{fig2B} \hfil }
   \centerline{\includegraphics[width=.9\columnwidth]{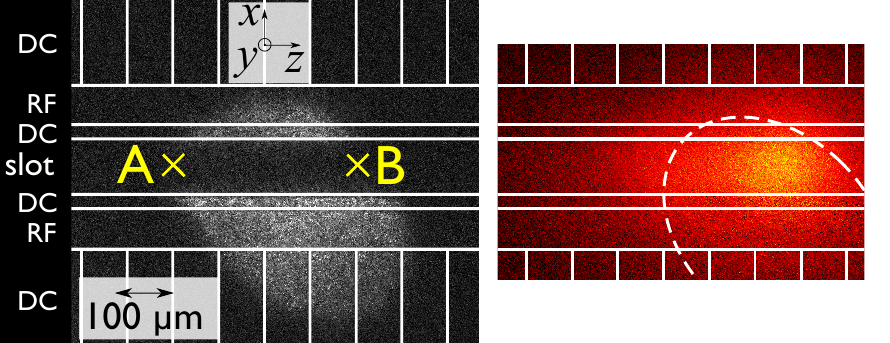}}
  \caption{Left: typical image of the cleaning laser light scattered from the trap. The electrode gaps are sketched as lines and the two trapping sites ``A'' and ``B''  discussed in the text are indicated.
Right: plume fluorescence (false colours) associated with a single cleaning pulse (energy density $\simeq 200$~mJ/cm$^2$) impinging on an uncleaned area.
The displayed image does not show the scattered laser light which was subtracted using an image taken after cleaning.}
  \label{fig:spots}
\end{figure}
%%%%%%%%
%
The cleaning beam is generated by a tripled Nd:YAG laser (Continuum Minilite ML I) that delivers 3--5~ns pulses (nominal) at $\lambda=355$~nm with an energy continuously adjustable up to $\sim 1$~mJ and a repetition rate of up to 15~Hz.
The beam is spectrally filtered by a fused silica prism and then sent to the trap.
A $\sim 300 \mu$m diameter pin-hole selects the central part of the laser beam and is imaged on the trap plane in order to obtain a well-defined spot with an intensity which is uniform to $\sim 20\%$. 
The imaging lens ($2f$--$2f$ configuration) is mounted on a micrometer translation stage to allow for fine positioning of the cleaning spot.
In view of the particular geometry of the trap, with a slot through the centre of the substrate, two symmetric beam paths, both at 45$^\circ$ incidence to the substrate plane ($xz$-plane), are used to allow cleaning of both interior walls of the slot (see figure~\ref{fig:scheme_trap}).   
The spot size and position on the trap are monitored using an EM-CCD camera (Andor Luca), also used for imaging of the trapped ion. 
A typical image of the UV light scattered from the trap is shown in figure~\ref{fig:spots} where the geometry of the electrodes is also sketched.
With this setup, energy densities up to $\simeq 350$~mJ/cm$^2$ can be obtained with a spot diameter of $\sim 300\mu$m.
In the following, energy densities are given normal to the beam propagation direction.
The energy density on a specific trap surface is reduced by a geometrical factor: for the $xz$-plane (upper electrode surfaces) this factor is $\frac{1}{\sqrt{2}}$, for the $yz$-plane (slot side walls) it is $\frac{1}{2}$.
All trap electrodes were grounded while firing the laser, to prevent the possibility of arcing initiated by photoelectrons.

%\section{Results}
%\label{sec:res}
%
%%%%%%%%
 \begin{figure}[h]
   \centerline{\includegraphics[width=.8\columnwidth]{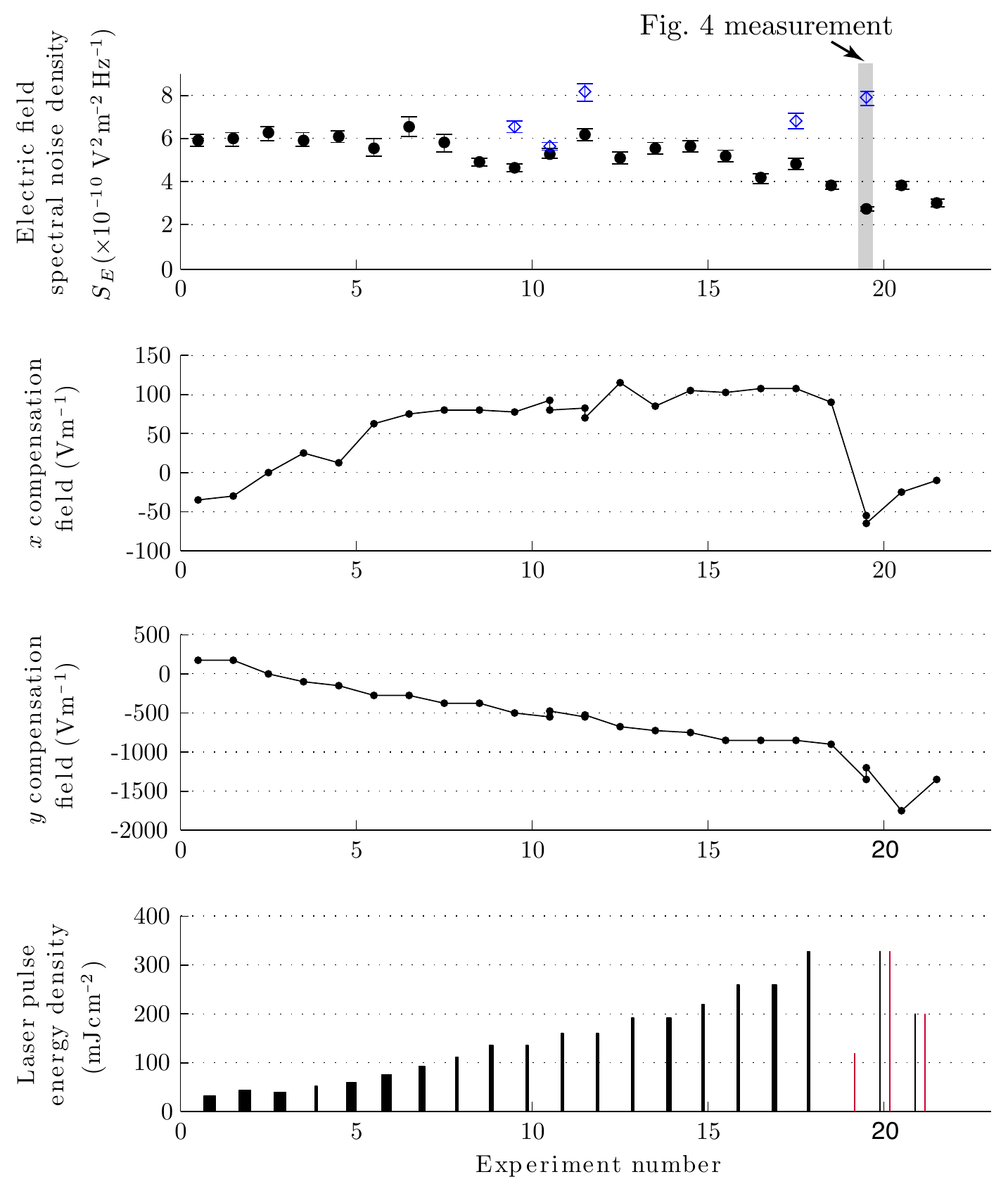}}
  \caption{Top to bottom: heating rates for an axial trapping frequency  $\omega_z/2\pi=500$~kHz (expressed in terms of electric field noise spectral density), micromotion compensation fields ($x$ and $y$ directions) and cleaning laser energy density plotted against the experiment number.
Each experiment consisted of a cleaning attempt followed by micromotion compensation and heating-rate measurements. 
In the top graph, blue open diamonds correspond to control measurements in site ``B'' while black filled circles correspond to measurements in the cleaned site ``A''.
Error bars are derived from the scatter in measurements repeated under nominally identical conditions.
In the bottom graph, the number of cleaning pulses applied in each experiment is proportional to the thickness of the bar (1000, 400 or 100 pulses); black (red) bars indicate cleaning from the $\vec{k_+}$  ($\vec{k_-}$) direction, respectively (see text). The entire data set was taken over a ten week period.}
 \label{fig:data}
\end{figure}
%%%%%%%
%
The experimental methods for loading and cooling $^{40}$Ca$^+$ ions in a similar trap, compensating micromotion,  and measuring heating rates using the Doppler re-cooling technique have previously been described in detail~\cite{Allcock:2010}.
Before applying any laser cleaning, we characterized the trap heating rate by testing three trapping sites ($z=0, \pm 240$~$\mu$m from centre).
The heating rate and frequency dependence were uniform (within the estimated error) and compatible with previous measurements~\cite{Allcock:2011}.
Contrary to the case of~\cite{Daniilidis:2011},  we did not observe an increase of the heating rate over an operation time of several months.
In our loading geometry the oven is placed below the slot shown in figure~\ref{fig:scheme_trap}, $\sim 50$~mm behind the trap.
Based on data from a similar oven~\cite{Lucas:2004}, we can estimate the order of magnitude of the flux reaching the  interior walls of the slot when the oven is on: $\sim 10^5$ Ca atoms per second and per mm$^2$ (less than 1 monolayer every 2 years).
However, we can not exclude a contamination of the slot surfaces by other species during the initial firing of the oven: at that moment, while the Ca flux was still $\lessapprox 10^7$ (atoms/s)/mm$^2$, the pressure in the vacuum chamber increased up to $10^{-8}$~mbar.

We began the laser cleaning by applying pulses to a trapping region two electrodes away from the centre of the trap array ($z=+160$~$\mu$m, site ``B'' on figure~\ref{fig:spots}).
Each experiment consisted of applying a number of pulses at a given energy density (1~Hz repetition rate) around the trapping position.
After each experiment the trap was loaded and the ion's heating rate was measured.
Then the energy density was increased for the next experiment.
At 30~mJ/cm$^2$ the DC electrode  centred at $x=-60$~$\mu$m showed clear signs of delamination near $z=+400$~$\mu$m, presumably caused by differential expansion induced by heating (see figure~\ref{fig:e-beam}a).
At this point no change in heating rates had been observed.
As we did not want to risk further damage to the trap we moved the trapping region to a symmetric position 4 electrodes away (site ``A'', $z=-160$~$\mu$m) and resumed the experiment, reducing the repetition rate to 0.2~Hz to minimize the risk of heat accumulation.
As we were still able to trap in site ``B'' and the heating rate had not significantly changed there, we later used it as a control measurement to ensure that any measured change in heating rate at zone ``A'' over time was not due to a systematic effect in our measurement or a change in some global noise source (e.g. electrical pickup).  
%
%%%%%%%%
 \begin{figure}[h]
   \centerline{\includegraphics[width=.75\columnwidth]{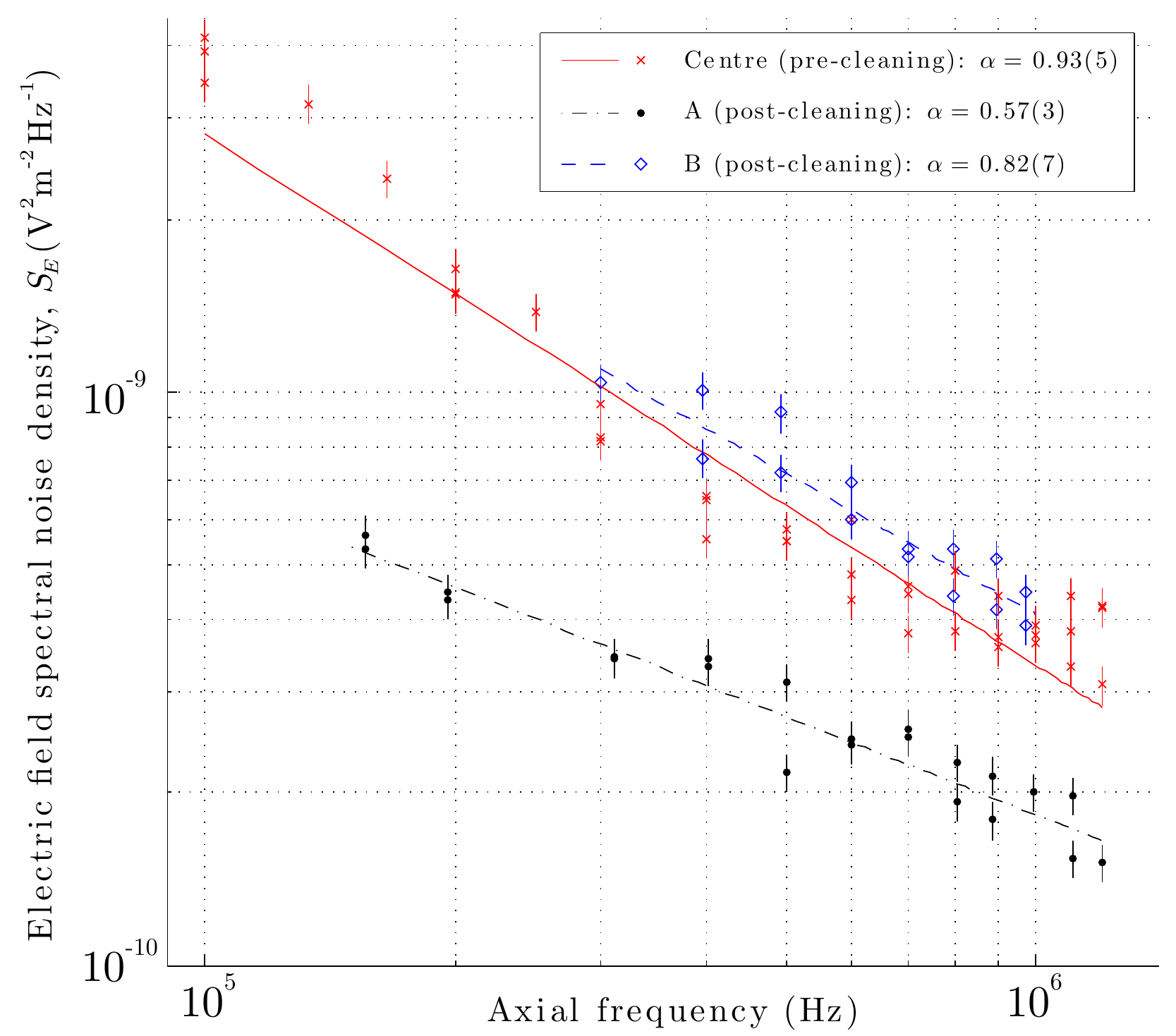}}
  \caption{Heating rate (expressed in terms of electric field noise spectral density) as a function of the axial trapping frequency $\omega_z/2\pi$.
We compare two data sets obtained in the ``A'' and ``B'' trapping sites (black filled circles  and blue open diamonds, respectively) taken on the same day with the same settings.
For reference, the data corresponding to the central site of the trap taken before cleaning (red crosses,~\cite{Allcock:2011}) are also shown.
Error bars are derived from the scatter in repeated measurement sets and the lines correspond to the best fits of each data set to a $\omega_z^{-\alpha}$ law.
The exponents $\alpha$ corresponding to the ``A'', ``B'' and ``centre'' sites are 0.57(3), 0.82(7) and 0.93(5) respectively.  
}
  \label{fig:ab}
\end{figure}
%%%%%%%
%

Figure \ref{fig:data}  shows the evolution of the heating rates (expressed in terms of electric field noise spectral density) and micromotion compensation fields throughout the entire series of cleaning experiments applied to site ``A''.
Initially we applied the cleaning beam along the $\vec{k_+}=(\frac{1}{2},-\frac{1}{\sqrt{2}},-\frac{1}{2})$ direction only (see figures~\ref{fig:scheme_trap} and \ref{fig:spots}), indicated by black bars in figure~\ref{fig:data}. 
There appears to be a slight drop in heating rates from $\sim100$~mJ/cm$^2$ onwards, which initially is not much below the scatter on the measurements.
However, once we attempted cleaning also from the $\vec{k_-}=(-\frac{1}{2},-\frac{1}{\sqrt{2}},\frac{1}{2})$ direction (indicated by red bars in figure~\ref{fig:data}) with an energy density of 100~mJ/cm$^2$, the drop became much more pronounced.
This effect points to a large contribution to the noise from the slot side wall, the only significant area not cleaned by the $\vec{k_+}$ directed beam.
When cleaning a ``fresh'' region, we also observed for each single-pulse shot a fluorescent emission (ablation plume) from inside the slot (see figure~\ref{fig:spots}) and an accompanying pressure spike of a few $10^{-12}$ Torr.
The plume fluorescence intensity and the pressure spike amplitude dropped rapidly and became undetectable after 3 or 4 shots, implying the source material responsible for these phenomena had been removed.
These effects were not observed in the first $\vec{k_+}$ cleaning direction.
This is due to the fact that along the $\vec{k_+}$ direction we gradually increased the intensity over thousands of pulses: this probably removed the material in smaller amounts, below the sensitivity of the camera or ion gauge.

At this point, heating rate data as a function of axial trap frequency was taken at both site ``A'' and ``B'' (see figure~\ref{fig:ab}).
Whilst the heating rate in site ``B'' is still entirely consistent with the data taken several months before~\cite{Allcock:2011}, that in site ``A'' shows a marked decrease and a significant drop of the exponent $\alpha$.

The exposure of the trap to cleaning laser pulses also caused a shift in the micromotion compensation voltages along both $x$ and $y$ directions.
These shifts had a small component ($\sim 10\%$) which relaxed over several hours (presumably induced by charging~\cite{Allcock:2011,Harlander:2010}) but the major part of the effect did not relax, even over weeks.
The direction of the electric field to be compensated was mainly such that the ion was attracted upwards ($+y$) and away from the side of the slot being cleaned.
The effect appeared to have somewhat saturated until the $\vec{k_-}$ cleaning direction was used, at which point the field roughly doubled in magnitude along $y$ but evened out in $x$.
Again, this behaviour points to a major contribution from the large slot side wall (silica or gold surfaces).

We then attempted to reduce the heating rate further by increasing the energy density to 360~mJ/cm$^2$ in the $\vec{k_-}$ direction; however this caused visible damage to the aluminium top surface of the trap (observed as an increase of light scattering in the irradiated zone).
This damage caused an increase in the heating rate (though still below the initial value) and a reversion to a higher exponent $\alpha$ in the frequency dependence [$\alpha= 0.88(3)$].

The still-operational microtrap was then removed from the vacuum chamber and observed under optical and electron microscopes.
Optical microscope images confirmed some visual damage of the Al surface of the electrodes surrounding site ``A''
and suggested a reflectivity decrease of the slot side wall where it had been irradiated.
Electron microscope images were taken both with secondary electron and back-scattered electron (BSE)  contrasts.
As shown in figure~\ref{fig:e-beam}~a, the delamination of the DC electrode (at $z\sim+400$~$\mu$m) appears to be associated with a delamination of the silica pillar, suggesting that some thermally-induced stress may be at the origin of this damage.
The boundary between irradiated (but not damaged) and non-irradiated region inside the slot side wall displays some change in the topography of the gold coating (figure~\ref{fig:e-beam}~b).
However, an image obtained with BSE contrast (which is sensitive to $Z$) shows that the gold was probably only removed (in a stripe-like fashion) around the damaged site ``A'' (figure~\ref{fig:e-beam}~c) and it still forms a continuous film in the regions irradiated with $< 200$~mJ/cm$^2$ energy density.

%\section{Discussion}
%\label{sec:discuss}

This study shows that the technique of high-intensity laser irradiation is capable of reducing {\it in situ } the heating rate of a microfabricated ion trap.
It is also notable that the exponent $\alpha$, characterizing the electric field noise frequency dependence, is affected by the procedure (figure~\ref{fig:ab}).

Two possible interpretations of the mechanism involved in this electric field noise reduction can be pointed out.
The first one is based on the (possibly partial) removal of surface contamination, responsible for the existence of patches~\cite{Daniilidis:2011,Safavi-Naini:2011}.
The theoretical study in~\cite{Safavi-Naini:2011} suggests that different adsorbates could give rise to different frequency dependences: partial cleaning of a sub-set of adsorbates could then explain the observed change in the exponent $\alpha$.
The second interpretation is that the observed effect was caused by the apparent change in the topography of the thin gold film inside the slot side wall (see figure~\ref{fig:e-beam}~b).
According to~\cite{Dubessy:2009, Low:2011}, this re-arrangement of the metallic film could increase the characteristic length $\zeta$ of the disorder, leading to a reduction of the heating-rate.
%
%%%%%%%%
 \begin{figure}[h]
  \centerline{\includegraphics[width=.55\columnwidth]{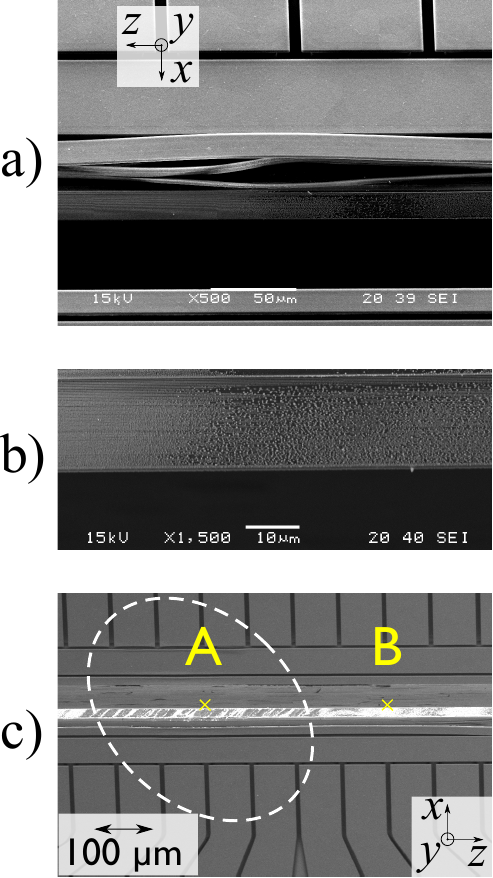}}
  \caption{Scanning electron microscope images of the microtrap (under a 45° angle to the $y$-axis, giving a view of a side wall of the slot) obtained after laser-cleaning experiments. a): delamination damage of DC electrode at $z=+400$~$\mu$m.
b): topography change between non-irradiated (left) and irradiated (right) zones of the side wall.
c): BSE contrast image of the side wall around the trapping site ``A''.
This side wall was irradiated with the maximum nominal energy density of $\simeq 360$~mJ/cm$^2$, i.e. $\simeq 180$~mJ/cm$^2$ after the geometrical correction.
Due to the $Z$-contrast of the BSE the gold shows up bright; it appears to have been completely removed in places. (the corresponding laser spot is sketched for reference)}
  \label{fig:e-beam}
\end{figure}
%%%%%%%
%
It should also be noted that in spite of the laser-cleaning procedure, we were unable to bring the measured heating rate below the best results obtained with traps of this size at room temperature ($S_E\sim 2\times10^{-12}$~V$^2$m$^{-2}$Hz$^{-1}$ at $\omega/2\pi=1$~MHz,  cf. figure~5 of~\cite{Daniilidis:2011}).
It is likely that surface contamination is only one contributing factor to anomalous heating.
If so, this technique could still be very useful as a method of reducing the large variance observed between traps of the same material and fabrication which currently renders any systematic studies into the best material and fabrication choice very difficult.

The particular microtrap that we used for this investigation was not ideally suited to the purpose; different materials were irradiated at the same time and the ion still had a direct line-of-sight to the dielectric pillars.
An improved version of the trap with shorter dielectric pillars and a front, as well as back, evaporated coating has already been demonstrated~\cite{Allcock:2011}.
A gold coating was used in that case but in principle any conducting material could be used.
A similar experiment with such a trap would be easier to interpret as only this one material predominates.

In order to develop this technique further a deeper understanding of cleaning and damage thresholds for typical trap structures and adsorbates is needed.
This could be done by combining laser-cleaning with analysis of the surface chemical composition (e.g. by Auger or X-ray photoelectron spectroscopy techniques).
Lasers with a higher photon energy (e.g. Nd:YAG 4th harmonic or excimer) or better ratio between peak intensity and average power (e.g. femtosecond lasers) should be investigated because more effective cleaning is expected for an equivalent thermal load.
Following these lines, an optimal combination of electrode materials and cleaning laser could be identified, allowing for routine {\it in situ} cleaning of microtraps whenever necessary. 

\ack
We are extremely grateful to Prof. P. Ewart and Dr. B. Williams for the loan of the Nd:YAG laser.
We thank D. Stick, D.L. Moehring, D. N. Stacey,  N. M. Linke and H. A. Janacek for helpful discussions.
L. G. thanks Balliol College in Oxford for an Oliver Smithies fellowship and acknowledges funding from EPSRC (grant EP/I028978/1).
This work was supported by a EPSRC Science \& Innovation Award.

%\bibliographystyle{unsrt}
%\bibliography{/Users/luca/Documents/biblio_total.bib}

\end{document}